\documentclass[pdflatex,sn-apa]{sn-jnl}

\usepackage{graphicx}%
\usepackage{multirow}%
\usepackage{amsmath,amssymb,amsfonts}%
\usepackage{amsthm}%
\usepackage{mathrsfs}%
\usepackage[title]{appendix}%
\usepackage{xcolor}%
\usepackage{textcomp}%
\usepackage{manyfoot}%
\usepackage{booktabs}%
\usepackage{algorithm}%
\usepackage{algorithmicx}%
\usepackage{algpseudocode}%
\usepackage{listings}%

\theoremstyle{thmstyleone}%
\theoremstyle{thmstyletwo}%
\newtheorem{remark}{Remark}%

\theoremstyle{thmstylethree}%

\begin{document}

\title[Article Title]{Extending Sheldon M. Ross’s Method for Efficient Large-Scale Variance Computation}
\author[ ]{\fnm{Jiawen} \sur{ Li}}\email{jiawen.li12@student.unsw.edu.au}

\affil[ ]{\orgdiv{School of Computer Science and Engineering}, \orgname{University of New South Wales}, \orgaddress{\street{Kensington}, \city{Sydney}, \postcode{2052}, \state{New South Wales}, \country{Australia}}}


\abstract{We introduce Prior Knowledge Acceleration (PKA), 
a batch-update method for variance that reuses previously 
computed sufficient statistics to avoid full recomputation. 
The update identity is algebraically equivalent to the 
pairwise formula of Chan, Golub, and LeVeque (1983); our 
contribution is a runtime-cost analysis that derives an 
explicit acceleration factor $\tau_a$ and identifies the 
data-size regime where batch updating outperforms both 
na\"ive recomputation and Ross's single-sample method. 
We prove that Ross's approach is preferable only when 
the new batch contains a single sample ($N_2 = 1$). 
We further generalise the framework to covariance and 
other decomposable statistics. Benchmarks against 
Welford, Chan pairwise, and na\"ive two-pass baselines 
on synthetic and real-world streaming data confirm the 
theoretical predictions, with speedups of up to 
$454\times$ when the prior dataset is large relative 
to the new batch.}


\maketitle

\section{Introduction and Related Works}
Variance plays a crucial role in data analysis, such as in ANOVA \cite{fisher}, and is widely applied in probability models within machine learning, including the Linear Gaussian Model \cite{gaussian} and Bayesian Regression. It is also used to estimate differences and contributions between models in ensemble learning \cite{ensemble}.

Despite its importance, computing variance and its variations for large datasets can be computationally expensive. Except for using various approaches to approximate the real value \cite{CT-Var}, or using matrix block computation to directly accelerate the computing process \cite{cov}, there still exists another solution to prior knowledge. However, beyond these computational strategies, leveraging prior knowledge presents another promising direction for optimization.

In the context of online machine learning, where models require continuous training and frequent parameter updates (e.g., gradients and loss values), variance estimation remains a challenge \cite{online}. These online machine learning models typically require constant training and parameter updates (such as gradients and loss values) with newly incoming data, and researchers have tried various methods like Expectation–maximization algorithm(EM) \cite{EM} or other numerical methods \cite{attempts} to estimate the variance \cite{promising}. Even though the ELM, refer as the Extreme Learning Machine(a type of model that could avoid frequent updates of parameters), or other alternative methods get proposed, the main issue in traditional online learning still remains \cite{ELM}. Hence PKA might be a way if get further extensions to covariance or directly adopt it when encountering similar tasks that require frequent parameter updates. 

The delay in online learning is always an issue that would usher low precision to the models, incur extra cost for operations, or add up the error within the calculation \cite{issue1} and even influence timely decision making \cite{issue2}. PKA could be one solution for it, as an instance, PKA could be seamlessly incorporated into the Mean Squared Error (MSE) calculation due to its similarity to variance. By adopting PKA in online machine learning, the risk of miscalculations caused by delays could be mitigated, leading to more accurate and efficient updates in real-time training. It is a feasible way since the result of PKA is an analytical solution albeit the potential limits.

\subsection{Related Work}\label{sec:related}

Efficient variance computation has been studied from two complementary
angles: \emph{numerical stability} and \emph{computational cost}.

\textbf{Single-sample online updates.}
Welford~\cite{Welford1962} introduced an iteration formula that derives
the corrected sum of squares $S_n$ for $n$ values from $S_{n-1}$ and the
new observation~$x_n$:
\begin{equation}\label{eq:welford}
  S_n = S_{n-1} + \frac{n-1}{n}(x_n - m_{n-1})^{2},
\end{equation}
where $m_n$ is the running mean.  Each value is used exactly once and
need not be stored, giving $O(1)$ auxiliary memory.  Welford also
extended the formula to corrected sums of products (covariance) and
higher-order moments.  The algorithms of West~\cite{West1979} and
Hanson~\cite{Hanson1975} are virtually identical to
Welford's~\cite{classic}.
Ross~\cite{ross} presented the same single-sample update in the
form of explicit expressions for the new mean $\bar{x}_{j+1}$ and
sample variance $s^2_{j+1}$, aimed at introductory statistics courses.

\textbf{Pairwise (parallel) updates.}
Youngs and Cramer~\cite{Youngs1971} gave an early updating formula
similar to Welford's.
Chan, Golub, and LeVeque~\cite{classic} generalised the single-sample
update to combine two sub-samples of \emph{arbitrary} sizes $m$ and~$n$.
Given the sums $T_{1,m}$, $T_{m+1,m+n}$ and the partial sums of squares
$S_{1,m}$, $S_{m+1,m+n}$, the combined sum of squares is
\begin{equation}\label{eq:chan}
  S_{1,m+n}
    = S_{1,m} + S_{m+1,m+n}
      + \frac{m}{n(m+n)}
        \Bigl(\tfrac{n}{m}\,T_{1,m} - T_{m+1,m+n}\Bigr)^{2}.
\end{equation}
Re-expressing~\eqref{eq:chan} in terms of sub-sample means and
population variances yields exactly the identity we derive
independently in our Eq.~(8).  The focus of Chan et
al.\ was on bounding the forward numerical error of each algorithm
variant; they provided a unified error-bound table (reproduced
conceptually in our Table~\ref{tab:comparison}) and a decision
procedure for choosing an algorithm based on the condition number
$\kappa$ of the data set.

\textbf{Extensions to higher-order moments and modern applications.}
P\'ebay~\cite{Pebay2008} extended the Chan et al.\ pairwise formula
to arbitrary-order centred moments and covariance, enabling one-pass
parallel computation for large-scale distributed data sets.
P\'ebay et al.~\cite{Pebay2016} further generalised these results
to multivariate moments with arbitrary weights and compound
moments, and provided a corrected two-pass algorithm that maintains
accuracy over nearly the full representable input range.
Bennett et al.~\cite{Bennett2009} demonstrated numerically stable
single-pass parallel statistics algorithms in high-performance
computing environments.
Schubert and Gertz~\cite{Schubert2018} presented a comprehensive
review that extends Welford's incremental technique to weighted
covariance and correlation, with applications to stock-market
analysis using exponentially weighted moving models and Gaussian
mixture modelling for cluster analysis.  Their work received the
SSDBM~2018 Best Paper Award and remains the most recent
comprehensive treatment.

All of the above works share a common concern for \emph{numerical
stability} and/or \emph{parallelism}.  None of them, however,
addresses the following questions:
\begin{enumerate}
  \item \emph{Runtime crossover.}
        For what combinations of prior dataset size~$N_1$ and new batch
        size~$N_2$ does the pairwise batch update become faster than
        full recomputation from scratch?  Chan et al.\ did not analyse
        the constant-factor computational cost, as their concern was
        numerical error rather than wall-clock time.
  \item \emph{Comparison with Ross's single-sample loop.}
        When $N_2 > 1$ new samples arrive, one may either apply Ross's
        (equivalently Welford's) update $N_2$ times or use the batch
        formula once.  No prior work has formally identified the
        crossover point.
  \item \emph{Generalisation to other decomposable statistics.}
        While the pairwise identity has been applied to variance and,
        implicitly, to higher moments, no unified framework has been
        proposed that captures the structural conditions under which
        batch updating is effective for an arbitrary decomposable
        function~$f$.
\end{enumerate}

\subsection{Our Contributions}\label{sec:contributions}

This paper addresses the three gaps identified above.  We call the
resulting family of methods \emph{Prior Knowledge Acceleration}
(PKA), since they leverage previously computed sufficient statistics
(the ``prior knowledge'') to avoid redundant computation.

\begin{enumerate}
  \item We re-derive the pairwise variance update---which is
        algebraically equivalent to Chan et al.'s Eq.~(1.5b)---and,
        for the first time, analyse it from a \emph{runtime-cost}
        perspective.  We introduce an explicit acceleration factor
        $\tau_a = t / t_p$ and derive the data-size regime in which
        $\tau_a > 1$, proving that batch updating strictly dominates
        full recomputation when $N_1$ is sufficiently large relative
        to~$N_2$ (Section~\ref{sec:proof}).

  \item We prove that Ross's single-sample update is preferable only
        when $N_2 = 1$; for all $N_2 \geq 2$ the batch formula has
        strictly lower operation count
        (Section~\ref{sec:ross_compare}).

  \item We extend the acceleration-factor analysis to a general
        decomposition framework applicable to any function of the
        form $f(D_1, D_2) = A\,f(D_1) + B\,f(D_2) + g(D_1, D_2)$,
        and demonstrate its utility on covariance as a worked example
        (Sections~\ref{sec:general}--\ref{sec:covariance}).

  \item We validate these predictions with reproducible benchmarks
        against Welford, Chan pairwise, and na\"ive two-pass baselines
        on both synthetic data and a real-world streaming dataset
        (Sections~\ref{sec:experiments}--\ref{sec:realtime}).
\end{enumerate}


\begin{table}[t]
\centering
\caption{Comparison of variance-update methods.  ``Prior state'' is the
  information retained between batches.  $N = N_1 + N_2$.}
\label{tab:comparison}
\small
\begin{tabular}{@{}lcccc@{}}
\toprule
\textbf{Method}
  & \textbf{Update cost}
  & \textbf{Prior state}
  & \textbf{Stability}
  & \textbf{Best when} \\
\midrule
Na\"ive two-pass
  & $\Theta(N)$
  & $O(N)$ (raw data)
  & Good
  & $D_1$ still available \\
Welford (1962)
  & $\Theta(N_2)$ per sample
  & $O(1)$: $(m, S, n)$
  & Good
  & Streaming, $N_2{=}1$ \\
Chan et al.\ (1983)
  & $\Theta(N_2)$ vectorised
  & $O(1)$: $(T, S, n)$
  & Good
  & Batch, large $N_1$ \\
\textbf{PKA (this paper)}
  & $\Theta(N_2)$ vectorised
  & $O(1)$: $(\sigma^2, \mu, n)$
  & Good
  & Same as Chan \\
Ross (2021) $\times N_2$
  & $\Theta(N_2)$ per sample
  & $O(1)$: $(\bar{x}, s^2, n)$
  & OK
  & Only $N_2{=}1$ \\
\bottomrule
\end{tabular}
\end{table}

\begin{figure}[H]
    \centering
    \includegraphics[width=0.75\linewidth]{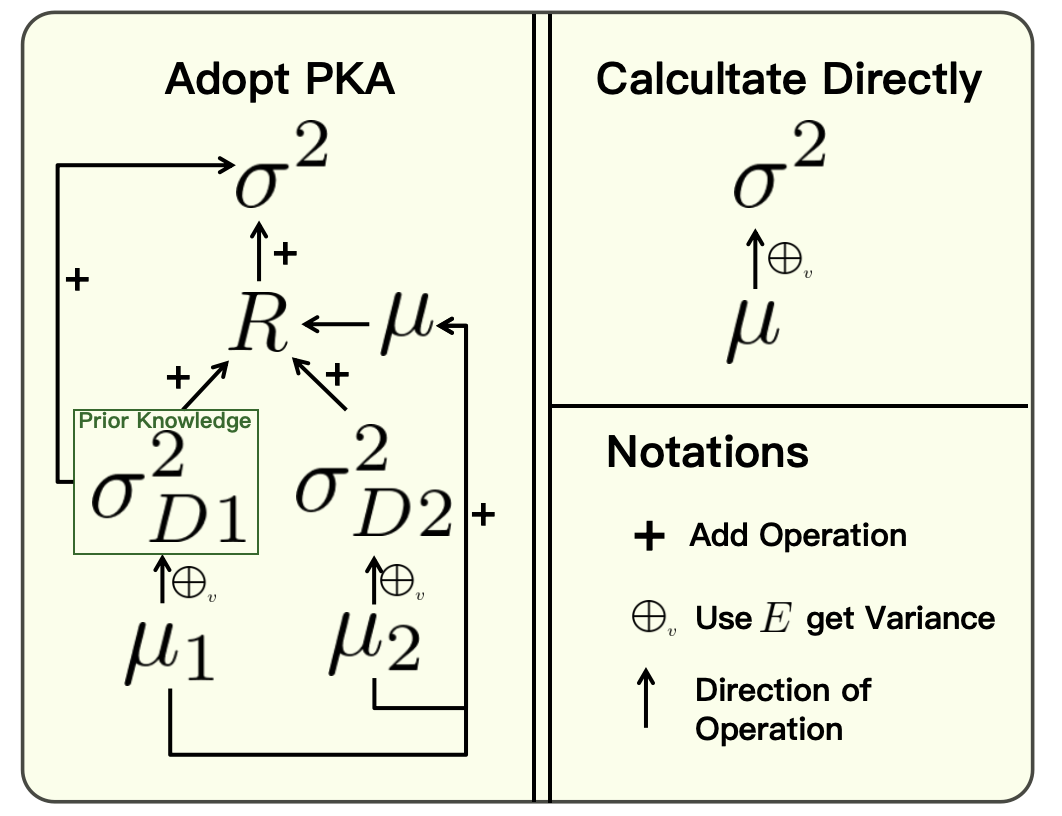}
    \caption{The Computing Graph for PKA}
    \label{fig:graph}
\end{figure}

\section{Methodology}
In this section, we derive the PKA update formulas for 
population and sample variance, analyse the conditions 
under which the update is faster than full recomputation, 
and compare PKA with Ross's single-sample method.

We call this family of methods Prior Knowledge Acceleration 
(PKA). While we focus primarily on variance, the framework 
extends naturally to covariance and other decomposable 
statistics, as demonstrated in Sections~\ref{sec:covariance}--\ref{sec:cov_proof}. The simplified overview of the computing graph for population variance as an example can be seen in Fig.\ref{fig:graph}.

\subsection{PKA(Prior Knowledge Acceleration)}
First, we defined a sequence $D$ representing the whole dataset after the extra data gets added to the original one in eq.\eqref{eq:1}. $D_1$ represent the original dataset, and $D_2$ is the added one. In addition, both of them are non-empty sets. Those sequences are not sets in the definition, which means we are not just simply defining $D_1$ and $D_2$ are independent to each other.
\begin{equation}
    D = (x_1, x_2, \dots, x_N), \quad D_1 = (x_1, x_2, \dots, x_{N_1}), \quad D_2 = (x_{N_1+1}, \dots, x_N)
    \label{eq:1}
    \tag{1}
\end{equation}

\subsubsection{PKA for Population Variance}
The eq.\eqref{eq:2} shows the main concept of this accelerating method, it assumes the variance of the whole dataset could be as formed in its original variance and an unknown remainder.We express the updated variance as the sum of the original variance and a "remainder" term, which accounts for the influence of the new data and the shift in the overall mean.
\begin{equation}
    \sigma_D^2=\sigma_{D_1}^2+R
    \label{eq:2}
    \tag{2}
\end{equation}

The paper defined the size of the data $D$ as $N$, and defined size $N_1$ and $N_2$ for its sub-sequences $D_1$, and $D_2$.
\begin{equation}
    N=N_1+N_2
    \label{eq:3}
    \tag{3}
\end{equation}

Getting rid of $\sigma_{D_1}^2$ in both side of eq.\eqref{eq:2}, it could transform to eq.\eqref{eq:4}.
\begin{equation}
    R \triangleq \sigma_D^2-\sigma_{D_1}^2
    \label{eq:4}
    \tag{4}
\end{equation}

We could extract the term $\sigma_{D_1}^2$ and $\sigma_{D_2}^2$ out of the expression from the definition(seen details from eq.\eqref{eq:5} to eq.\eqref{eq:7}.) of the population variance $\sigma^2$ of $D$, and then we could get the simplified result in eq.\eqref{eq:8}.

\begin{equation}
    \sigma_{D}^2=\frac{\sum_{x_i \in D} (x_i-\mu)^2}{N}
    \label{eq:5}
    \tag{5}
\end{equation}

\begin{equation}
    \sigma_{D}^2=\frac{1}{N}(\sum_{x_j \in D_1} (x_j-\mu)^2+\sum_{x_k \in D_2} (x_k-\mu)^2)
    \label{eq:6}
    \tag{6}
\end{equation}

\begin{equation}
    \sigma_{D}^2=\frac{1}{N}(\sum_{x_j \in D_1} (x_j-\mu_1+\mu_1-\mu)^2+\sum_{x_k \in D_2} (x_k-\mu_2+\mu_2-\mu)^2)
    \label{eq:7}
    \tag{7}
\end{equation}

\begin{equation}
    \sigma_{D}^2=\frac{1}{N}(N_1\sigma_{D_1}^2+N_2\sigma_{D_2}^2+N_1(\mu_1-\mu)^2+N_2(\mu_2-\mu)^2)
    \label{eq:8}
    \tag{8}
\end{equation}

Note that Eq.~\eqref{eq:8} is algebraically equivalent to 
the pairwise update of Chan et al.~\cite{classic}. Our 
contribution is not the identity itself, but the runtime-cost analysis in Section~\ref{sec:proof} and the 
generalisation in Section~\ref{sec:general}.

Since we need to get the expression of the Remainder $R$, we could put back the eq.\eqref{eq:8} into eq.\eqref{eq:2}. Hence, the final form of the Remainder could explicate in beneath(seen in eq.\eqref{eq:10}):

\begin{equation}
    R=\frac{1}{N}(N_1\sigma_{D_1}^2+N_2\sigma_{D_2}^2+N_1(\mu_1-\mu)^2+N_2(\mu_2-\mu)^2)-\sigma_{D_1}^2
    \label{eq:9}
    \tag{9}
\end{equation}

\begin{equation}
    R=\frac{1}{N}(N_2(\sigma_{D_2}^2-\sigma_{D_1}^2)+N_1(\mu_1-\mu)^2+N_2(\mu_2-\mu)^2)
    \label{eq:10}
    \tag{10}
\end{equation}

As for the mean value of PKA in variance(it had been widely use due to its simplicity, it's a special case of PKA that do not has the third term) when knowing $\mu_1$ and $\mu_2$, the expression could seen in eq.\eqref{eq:11}. Therefore, we could use the R add with the original population variance to get the updated one.
\begin{equation}
    \mu=\frac{N_1\mu1+N_2\mu2}{N}
    \label{eq:11}
    \tag{11}
\end{equation}

\subsubsection{PKA for Sample Variance}
What if we are calculating sample variance instead of the population one? The method still follows the same structure as in the population variance case, but with slight adjustments to it.We could get back to eq.\eqref{eq:7} and substitute the variable in it to get eq.\eqref{eq:12}.

\begin{equation}
    S_{D}^2=\frac{1}{N-1}(\sum_{x_j \in D_1} (x_j-\Bar{X_j}+\Bar{X_j}-\Bar{X})^2+\sum_{x_k \in D_2} (x_k-\Bar{X_k}+\Bar{X_k}-\Bar{X})^2)
    \label{eq:12}
    \tag{12}
\end{equation}

\begin{equation}
    =\frac{1}{N-1}((N_1-1)S_{D_1}^2+(N_2-1) S_{D_2}^2+N_1 (\Bar{X_j}-\Bar{X})^2+N_2 (\Bar{X_k}-\Bar{X})^2)
    \label{eq:13}
    \tag{13}
\end{equation}

Using a similar process of eq.\eqref{eq:9} in eq.\eqref{eq:13}, consequently, we could finally get the expression of $R$ in eq.\eqref{eq:15}. 

\begin{equation}
    R=\frac{1}{N-1}((N_1-1)S_{D_1}^2+(N_2-1) S_{D_2}^2+N_1 (\Bar{X_j}-\Bar{X})^2+N_2 (\Bar{X_k}-\Bar{X})^2)-S_{D_1}^2
    \label{eq:14}
    \tag{14}
\end{equation}

\begin{equation}
    R=\frac{1}{N-1}((N_2-1) S_{D_2}^2-N_2 S_{D_1}^2)+N_1 (\Bar{X_j}-\Bar{X})^2+N_2 (\Bar{X_k}-\Bar{X})^2)
    \label{eq:15}
    \tag{15}
\end{equation}

\subsubsection{Proof of PKA's effectiveness in Population Variance}
\label{sec:proof}
\begin{algorithm}[h]
\caption{PKA Variance Update}\label{alg:pka}
\begin{algorithmic}[1]
\Require Prior statistics $(\mu_1, \sigma^2_{D_1}, N_1)$; new batch $D_2$
\Ensure Updated variance $\sigma^2_D$, mean $\mu$, count $N$
\State $N_2 \gets |D_2|$
\State $N \gets N_1 + N_2$
\State $\mu_2 \gets \frac{1}{N_2}\sum_{x \in D_2} x$
\State $\sigma^2_{D_2} \gets \frac{1}{N_2}\sum_{x \in D_2}(x - \mu_2)^2$
\State $\mu \gets \frac{N_1 \mu_1 + N_2 \mu_2}{N}$
\State $R \gets \frac{1}{N}\big(N_2(\sigma^2_{D_2} - \sigma^2_{D_1}) 
        + N_1(\mu_1 - \mu)^2 + N_2(\mu_2 - \mu)^2\big)$
\State $\sigma^2_D \gets \sigma^2_{D_1} + R$
\State \Return $(\mu,\; \sigma^2_D,\; N)$
\end{algorithmic}
\end{algorithm}

\paragraph{Complexity.}
Given prior sufficient statistics $(\mu_1, \sigma^2_{D_1}, N_1)$, 
Algorithm~\ref{alg:pka} processes the new batch $D_2$ in 
$\Theta(N_2)$ time and $O(1)$ additional stored state (three 
scalars). By contrast, na\"ive recomputation from the concatenated 
data requires $\Theta(N_1 + N_2)$ time and $O(N_1 + N_2)$ space, 
since the entire dataset must be available. Welford's online 
update~\cite{Welford1962} also uses $O(1)$ state but processes 
each sample individually, requiring $\Theta(N_2)$ scalar 
iterations with higher per-element constant cost than the 
vectorised operations in Algorithm~\ref{alg:pka}. Ross's 
single-sample formula~\cite{ross} applied $N_2$ times likewise 
costs $\Theta(N_2)$ scalar iterations; as shown in 
Section~\ref{sec:ross_compare}, its total operation count exceeds PKA's 
for all $N_2 \geq 2$.

In the paper, we assume the computation time as a unit for one single addition is $u_a$, and $u_m$ for one single multiplication. $M$ is the number of multiplications needed to compute, and $A$ is the number of additions. Based on the definition of Table \ref{tab:PKA1}, we get the computing time $t_{p}$ of $\sigma_{D}^2$ in PKA, it could be written as a form in eq.\eqref{eq:16}.

\begin{table}[h]
\caption{Computational Times for Variance}
\begin{tabular}{p{1.8cm}|p{1.8cm}p{1.8cm}}
\toprule
Term&A&M\\
\midrule\
$\sigma_{D}^2$&2{N}-1&N\\
$\sigma_{D_1}^2$&2$N_1$-1&$N_1$\\
$\sigma_{D_2}^2$&2$N_2$-1&$N_2$\\
\midrule
$\mu$&$N-1$&1\\
$\mu$ in PKA&1&3\\
$\mu_1$&$N_1-1$&1\\
$\mu_2$&$N_2-1$&1\\
\midrule
$R$&5&6\\
$R$ in $S_{D}^2$&7&6\\
Total $R$&$N+2N_2+3$&$N_2+11$\\
$\sigma_{D}^2$ in PKA&$N+2N_2+3$&$N_2+12$\\
$S_{D}^2$ in PKA&$N+2N_2+5$&$N_2+12$\\
\bottomrule
\end{tabular}
\label{tab:PKA1}
\end{table}

\begin{equation}
    t_{p}=u_a(N+2N_2+3)+u_m(N_2+13)
    \label{eq:16}
    \tag{16}
\end{equation}

Next step, we need to compare the relation between the $t_p$ and the computing time of calculating variance $t$ directly. 
\begin{equation}
    t=u_a(2N-1)+u_m N
    \label{eq:17}
    \tag{17}
\end{equation}

When PKA is effectiveness means $t-t_{p}>0$, we further simplified it and defined an accelerating factor $\tau_a$, the factor gets smaller means PKA is more effective in calculating variance.

\begin{equation}
    \tau_a \triangleq \frac{t}{t_{p}}>1
    \label{eq:18}
    \tag{18}
\end{equation}

Considering the extreme conditions, when $N_1$ approaches infinity, which means we hold an extremely large original dataset in that case, the limit of factor $\tau_a$ is:
\begin{equation}
    \tau_a=\lim_{N_1 \to \infty} \frac{t}{t_{p}}=\lim_{N_1 \to \infty} \frac{u_a(2N-1)+u_m N}{u_a(N+2N_2+3)+u_m(N_2+12)}=\frac{2u_a+u_m}{u_a}>1
    \label{eq:19}
    \tag{19}
\end{equation}

But as for increasing $N_2$, the factor is less than the one illustrated in eq.\eqref{eq:20}, which is a condition that PKA does nothing or worse than directly calculating the definition. This explains that, compared to the size of added data, the PKA method is only established and effective with enough original data. 

\begin{equation}
    \tau_a=\lim_{N_2 \to \infty} \frac{t}{t_{p}}=\lim_{N_2 \to \infty} \frac{u_a(2N-1)+u_m N}{u_a(N+2N_2+3)+u_m(N_2+12)}=\frac{2u_a+u_m}{3u_a+u_m}<1
    \label{eq:20}
    \tag{20}
\end{equation}

\subsection{PKA compare with Sheldon M. Ross's Method}
\label{sec:ross_compare}
By transforming the definition of sample variance, Sheldon M. Ross suggests an approach for calculating the new variance and mean value after adding one new sample into the data, representing in eq.\eqref{eq:21} as well as eq.\eqref{eq:22}  \cite{ross}.(This section using original signs in the book for fast references)

\begin{equation}
    \Bar{x}_{j+1}=x_{j}-\frac{\Bar{x}_{j}}{j+1}
    \label{eq:21}
    \tag{21}
\end{equation}

\begin{equation}
    s_{j+1}^2=(1-\frac{1}{j})s_{j}^2+(j+1)(\Bar{x}_{j+1}-\Bar{x}_{j})^2
    \label{eq:22}
    \tag{22}
\end{equation}

According to the definition, we could get $t_r$, the time of the Sheldon M. Ross's Method when adding $N_2$ samples with time function $T$ in eq.\eqref{eq:24}:

\begin{equation}
    t_r=T(\Bar{x}_{N_1})+N_2(T(\Bar{x}_{j+1})+T(s_{j+1}^2))
    \label{eq:23}
    \tag{23}
\end{equation}

\begin{equation}
    T(\Bar{x}_{N_1})=(N_1-1)u_a+u_m,
    T(\Bar{x}_{j+1})=2u_a+u_m,
    T(s_{j+1}^2)=4u_a+3u_m
    \label{eq:24}
    \tag{24}
\end{equation}

\begin{equation}
    t_r=(N+5N_2-1)u_a+(4N_2-1)u_m
    \label{eq:25}
    \tag{25}
\end{equation}

\begin{equation}
    t_p'=t_p+(T(R')-T(R))=t_p+2u_a=u_a(N+2N_2+5)+u_m(N_2+13)
    \label{eq:26}
    \tag{26}
\end{equation}Further comparing the $t_r$ and $t_p'$(the running of PKA in sample variance) in eq.\eqref{eq:25} and eq.\eqref{eq:26}, we could see that $t_r$ only hold a smaller computing time of additions when $N_2<2$(which means $N_2$ can only be one), but this equation conflicts with the scenario that $N_2$ have to be greater or equal to 2 for having variance. As for multiplications, $t_r<t_p'$ established only when $N_2<int(\frac{11}{3})=3$. To summarize those two conditions to ensure $t_r$ must be smaller than $t_p'$, $N_2$ has to equal to one. In conclusion, for dealing with an extremely small amount of adding data, directly Sheldon M. Ross's method may present better than PKA in Variance, otherwise it not be a good choice. 

\subsection{General Form of PKA}
\label{sec:general}

\begin{remark}
The following discussion provides an informal framework 
for understanding when PKA-type decompositions are 
beneficial. The concrete effectiveness conditions for 
variance (Section~2.1.3) and covariance (Section~2.4.2) 
remain fully rigorous.
\end{remark}

After discussing PKA in computing variance, we could further consider how this sort of approach is applied on a more general scale. In the general form of PKA, it defines two datasets or vectors $D_1$ and $D_2$, and holds other same assumptions. In general PKA, the distinction is that here we defined a function $f$, which is the function the task aims to calculate rather than variance, and also establish a function $g$ to represent what the paper called remainder when calculating variance:
\begin{equation}
    \forall D_1,D_2,f(D_1,D_2)\triangleq A f(D_1)+B f(D_2)+g(D_1,D_2)
    \label{eq:27}
    \tag{27}
\end{equation}

Once we know the prior knowledge $D_1$, it become constant values, which we using $C1$ to distinguish them:
\begin{equation}
    \because f(C1,D_2) \triangleq A f(C1)+B f(D_2)+g(C1,D_2)
    \label{eq:28}
    \tag{28}
\end{equation}

When the PKA has a smaller computing time, it simply means $T(f(D_1,D_2))>T(f(C1,D_2))$. Additionally, the time function $T$ is a linear function that is a linear combination of unit time. The $Bf(D_2)$ term get offset, hence further conduct we can get:
\begin{equation}
    \therefore A\times(T(f(D_1))-T(f(C1)))>T(g(C1,D_2))-T(g(D_1,D_2))
    \label{eq:29}
    \tag{29}
\end{equation}

Due to there being no demand to calculate $C1$, the term in eq.\eqref{eq:30} must be positive. 
\begin{equation}
    \because T(f(D_1))-T(f(C1))>0
    \label{eq:30}
    \tag{30}
\end{equation}

So we could divide the right side in both sides to get the general form PKA factor $\tau_a$ in eq.\eqref{eq:31}. The condition is exactly the same as the PKA in variance, accelerate factor needs to be larger than 1.
\begin{equation}
     \tau_a \triangleq A\frac{T(f(D_1))-T(f(C1))}{T(g(C1,D_2))-T(g(D_1,D_2))}
    \label{eq:31}
    \tag{31}
\end{equation}

To have a deeper understanding of this condition, we could construct two new function called $z$ and $h$, and we treat $D_2$ as constant in the $h$ function, the definition of it could seen in beneath:
\begin{equation}
    z(x)\triangleq (T\circ f)(x), h(z)\triangleq(g\circ f^{-1})(T^{-1}(z),D_2)
    \label{eq:32}
    \tag{32}
\end{equation}

In that case, eq.\eqref{eq:29} could convert to:
\begin{equation}
    \therefore z(D_1)-z(C1)>(h\circ z)(C1)-(h\circ z)(D_1)
    \label{eq:33}
    \tag{33}
\end{equation}

The left side of the inequality is larger than the right, so multiplying both sides by a positive constant $L$, and in here is L=1, and taking the absolute value preserves the inequality in eq.\eqref{eq:34}. This satisfies Lipschitz's condition  \cite{lip}. Intuitively, the inequality suggests that the computational savings from caching $f(D_1)$ outpace the additional cost of evaluating~$g$, providing a practical condition for PKA's effectiveness. When a constraint limits $D_1$ in a finite dataset, these conditions guarantee the existence of extreme values. $h(x)$ describes the relationship between the original and accelerated calculation times, showing that PKA works in general situations.

\begin{equation}
    \therefore L|z(D_1)-z(C1)|>|(h\circ z)(C1)-(h\circ z)(D_1)|
    \label{eq:34}
    \tag{34}
\end{equation}

However, when the function values of the entire dataset involve recursive relationships, as seen in the Master Method, the PKA method becomes inapplicable. Similarly, ill-functions like oscillations or discontinuities may make the PKA method ineffective, or the function highly out of the linearity(like transcendental functions), or yet the condition about $\tau$ gets satisfied but the size of data does not reach the range where certainly holds extreme values. But in most cases, PKA still could guarantee its acceleration.

\subsection{Examples of using PKA}
Excepting the variance or mean value(in eq.\eqref{eq:11}, also a special case of using PKA methods but with $g(D_1, D_2))=0$) we discussed, there are other functions that could decompose in that way for faster computation including covariance, or other resemble instances like Within-Cluster Sum of Squares(WCSS) in K-Means Clustering\cite{WCSS}, Sum of Square in ANOVA, etc, also could adopt it. This section of the paper would further illustrate a deeper understanding of how to utilize it instead of merely variance, using PKA in covariance as an example.

\subsubsection{PKA for Covariance}
\label{sec:covariance}
The pairwise covariance update has been derived by 
P\'ebay~\cite{Pebay2008}; we re-derive it here within 
our PKA framework to demonstrate the generality of 
the decomposition in Eq.~\eqref{eq:27}.

Covariance is the more general form of regular variance, it is a good example to show PKA's utility. According to the definition of covariance, the 2 dimensional covariance of the whole dataset $\text{Cov}_D$ could be written as in eq.\eqref{eq:35}:
\begin{equation}
    \text{Cov}_D = \frac{1}{N} \sum_{i \in D} (x_i - \mu_{x,D})(y_i - \mu_{y,D})
    \label{eq:35}
    \tag{35}
\end{equation}

Similarly, we first decompose this $\text{Cov}_D$ into two expressions, considering two realms of sum $D_1$,$D_2$, and multiply N on both sides. In that case, we could obtain equation \eqref {eq:36}. (About $\mu$, the first subscript represents the variable of mean value belonging to, and the second one means which dataset we discuss. Ex:$\mu_{y,1}$ is the mean value of y in dataset $D_1$)
\begin{equation}
     \sum_{i \in D_1} (x_i - \mu_{x,D})(y_i - \mu_{y,D}) + \sum_{i \in D_2} (x_i - \mu_{x,D})(y_i - \mu_{y,D})
    \label{eq:36}
    \tag{36}
\end{equation}

Now, we turn to focus on the first term about $D_1$, we add a new term and minus it, which is mathematically equivalent, hence we could get the new forms of $(x_i - \mu_{x, D})$ as well as $(y_i - \mu_{y, D})$ in eq.\eqref{eq:37}\eqref{eq:38}:
\begin{equation}
     x_i - \mu_{x,D} = (x_i - \mu_{x,1}) + (\mu_{x,1} - \mu_{x,D})
    \label{eq:37}
    \tag{37}
\end{equation}
\begin{equation}
     y_i - \mu_{y,D} = (y_i - \mu_{y,1}) + (\mu_{y,1} - \mu_{y,D})
    \label{eq:38}
    \tag{38}
\end{equation}

Substituting eq.\eqref{eq:37}\eqref{eq:38} back into eq.\eqref{eq:36}, it transform into:
\begin{equation}
\sum_{i \in D_1}  \big( (x_i - \mu_{x,1}) + (\mu_{x,1} - \mu_{x,D}) \big) \cdot \big( (y_i - \mu_{y,1}) + (\mu_{y,1} - \mu_{y,D})
    \label{eq:39.1}
    \tag{39.1}
\end{equation}

Expanding eq.\eqref{eq:39.1} gives four parts:
\begin{equation}
= \sum_{i \in D_1} (x_i - \mu_{x,1})(y_i - \mu_{y,1})
\label{eq:39.2}
\tag{39.2}
\end{equation}
\begin{equation}
+ \sum_{i \in D_1} (x_i - \mu_{x,1})(\mu_{y,1} - \mu_{y,D})
\label{eq:39.3}
\tag{39.3}
\end{equation}
\begin{equation}
+ \sum_{i \in D_1} (\mu_{x,1} - \mu_{x,D})(y_i - \mu_{y,1})
\label{eq:39.4}
\tag{39.4}
\end{equation}
\begin{equation}
+ \sum_{i \in D_1} (\mu_{x,1} - \mu_{x,D})(\mu_{y,1} - \mu_{y,D})
\label{eq:39.5}
\tag{39.5}
\end{equation}

Now, in eq.\eqref{eq:39.2}, we could see the first term is exactly the definition of $N_1$ times of $Cov_1$(covariance of dataset1). As for eq.\eqref{eq:39.3}, since term $(\mu_{y,1} - \mu_{y, D})$ is a constant, it could get out of sum, and left $\sum_{i \in D_1} (x_i - \mu_{x,1})$, which is zero based on the definition of mean value, and that is the same for eq.\eqref{eq:39.5}.

In summary, the form of the first sum about $D_1$ is:
\begin{equation}
\sum_{i \in D_1} (x_i - \mu_{x,D})(y_i - \mu_{y,D}) = N_1 \cdot \text{Cov}_1 + N_1 (\mu_{x,1} - \mu_{x,D})(\mu_{y,1} - \mu_{y,D})
\label{eq:40}
\tag{40}
\end{equation}

Vice versa, get the form of the sum term for $D_2$, and merge it with the first one:
\begin{equation}
N \cdot \text{Cov}_D =  \big( N_1 \cdot \text{Cov}_1 + N_1 (\mu_{x,1} - \mu_{x,D})(\mu_{y,1} - \mu_{y,D}) \big) 
\label{eq:41.1}
\tag{41.1}
\end{equation}
\begin{equation}
+ \big( N_2 \cdot \text{Cov}_2 + N_2 (\mu_{x,2} - \mu_{x,D})(\mu_{y,2} - \mu_{y,D}) \big)
\label{eq:41.2}
\tag{41.2}
\end{equation}

Organizing the expression, and writing it into general form eq.\eqref{eq:27}, we could finally get the corresponding value of each term:
\begin{equation}
f(D) = \text{Cov}_D, f(D1) = \text{Cov}_1, f(D2) = \text{Cov}_2
\label{eq:42.1}
\tag{42.1}
\end{equation}
\begin{equation}
A = \frac{N_1}{N}, B = \frac{N_2}{N}
\label{eq:42.2}
\tag{42.2}
\end{equation}
\begin{equation}
g(D1, D2) = \frac{N_1}{N}(\mu_{x,1} - \mu_{x,D})(\mu_{y,1} - \mu_{y,D}) + \frac{N_2}{N}(\mu_{x,2} - \mu_{x,D})(\mu_{y,2} - \mu_{y,D})
\label{eq:42.3}
\tag{42.3}
\end{equation}
After writing covariance in the form of PKA, we could further discuss its effectiveness as we did in variance.

\subsubsection{Proof of Effectiveness of PKA for Covariance}
\label{sec:cov_proof}
Have a recap of the computational time in Table \ref{tab:PKA1}, we were treating the time of computing all could be decomposed by multiples (also including divisions) and additions (including subtraction), based on the previous formula in eq.\eqref{eq:42.1}\eqref{eq:42.2}\eqref{eq:42.3} the time of using PKA in covariance is:

\begin{table}[h]
\centering
\caption{Computational Times for Covariance}
\begin{tabular}{@{}l|cc@{}}
\toprule
\textbf{Term} & \textbf{A} & \textbf{M} \\
\midrule
$\text{Cov}_D$ (Baseline) & $5N - 3$ & $N + 3$ \\
\midrule
Stats for $D_2$ & $5N_2 - 3$ & $N_2 + 3$ \\
Global Mean Update & $2$ & $6$ \\
Covariance Update Step & $7$ & $8$ \\
\midrule
$\text{Cov}_D$ (PKA Total) & $5N_2 + 6$ & $N_2 + 17$ \\
\bottomrule
\end{tabular}
\label{tab:PKA2}
\end{table}

In Table \ref{tab:PKA2} we could write the total computational time of baseline(direct calculate covariance) $t_{cov}$ and PKA $t_{covp}$ into:

\begin{equation}
t_{cov}=(5N - 3)u_a+(N_2 + 17)u_m
\label{eq:43}
\tag{43}
\end{equation}

\begin{equation}
t_{covp}=(5N_2 + 6)u_a+(N + 3)u_m
\label{eq:44}
\tag{44}
\end{equation}

To sum up,in PKA method, the factor $\tau_{a}$ for covariance is:
\begin{equation}
\tau_{a}=\frac{t}{t_p}=\frac{t_{cov}}{t_{covp}}=\frac{(5N - 3)u_a+(N_2 + 17)u_m}{(5N_2 + 6)u_a+(N + 3)u_m}
\label{eq:45}
\tag{45}
\end{equation}

To fulfill the condition in eq.\eqref{eq:18}, it have to be:
\begin{equation}
\frac{(5N - 3)u_a+(N_2 + 17)u_m}{(5N_2 + 6)u_a+(N + 3)u_m}>1
\label{eq:46}
\tag{46}
\end{equation}

Reorganizing eq.\eqref{eq:46}, it could write as eq.\eqref{eq:47},and finally could yield the condition for effectiveness about $N_1$ in eq.\eqref{eq:48}:
\begin{equation}
(5N_1 + 3)u_a+(-N_1 + 14)u_m>0
\label{eq:47}
\tag{47}
\end{equation}

\begin{equation}
N_1>-\frac{3u_a+14u_m}{5u_a-u_m}
\label{eq:48}
\tag{48}
\end{equation}

When $5u_a-u_m>=0$, the condition is always standing since $N_1>0$. When $5u_a-u_m<0$, which means $5u_a<u_m$. Based on that, we could change the $3u_a+14u_m$ into $15u_m$, which is surely larger than the original term. For the denominator, we could minus $5u_a$, leading the whole expression become eq.\eqref{eq:49}:
\begin{equation}
N_1>15=-\frac{15u_m}{-u_m}>-\frac{3u_a+14u_m}{5u_a-u_m}
\label{eq:49}
\tag{49}
\end{equation}

\section{Simulated Tests of PKA in Population Variance}
\label{sec:experiments}

\paragraph{Implementation note.}
Chan pairwise and PKA are implemented with NumPy 
vectorised operations, while Welford and Ross use 
per-sample Python loops. This reflects the practical 
batch-update scenario where data arrives as arrays. 
A pure-C comparison would narrow the gap but not 
eliminate it, since batch methods perform $O(1)$ 
passes over $D_2$ versus $O(N_2)$ scalar updates.

As for validating that the method is truly accelerating and has its value, the experiment is settled in a standard Kaggle environment, making it easy to replicate. By utilizing \textbf{Numpy} package in the \textbf{first test}, we generate 25,0000 random samples in $D_1$ and 25,0000 random samples in $D_2$, and those generated data all follow uniform distribution(because the distributions not effect the computational time of variance). 

When $N_1=N_2$ is in the same situation as $N_1$ and $N_2$ both approximate positive infinity and it's the case that the PKA has better performance compared to the baseline. The paper taking the mean values of 10,0000 running time of one single operation and knowing that the Kaggle environment holds approximately $\tau_a=1.2080$ with $u_a=2.2238-07, u_m=2.3384e-07$, which satisfies the condition $\tau_a>1$.

\begin{figure}[H]
    \centering
    \includegraphics[width=0.8\linewidth]{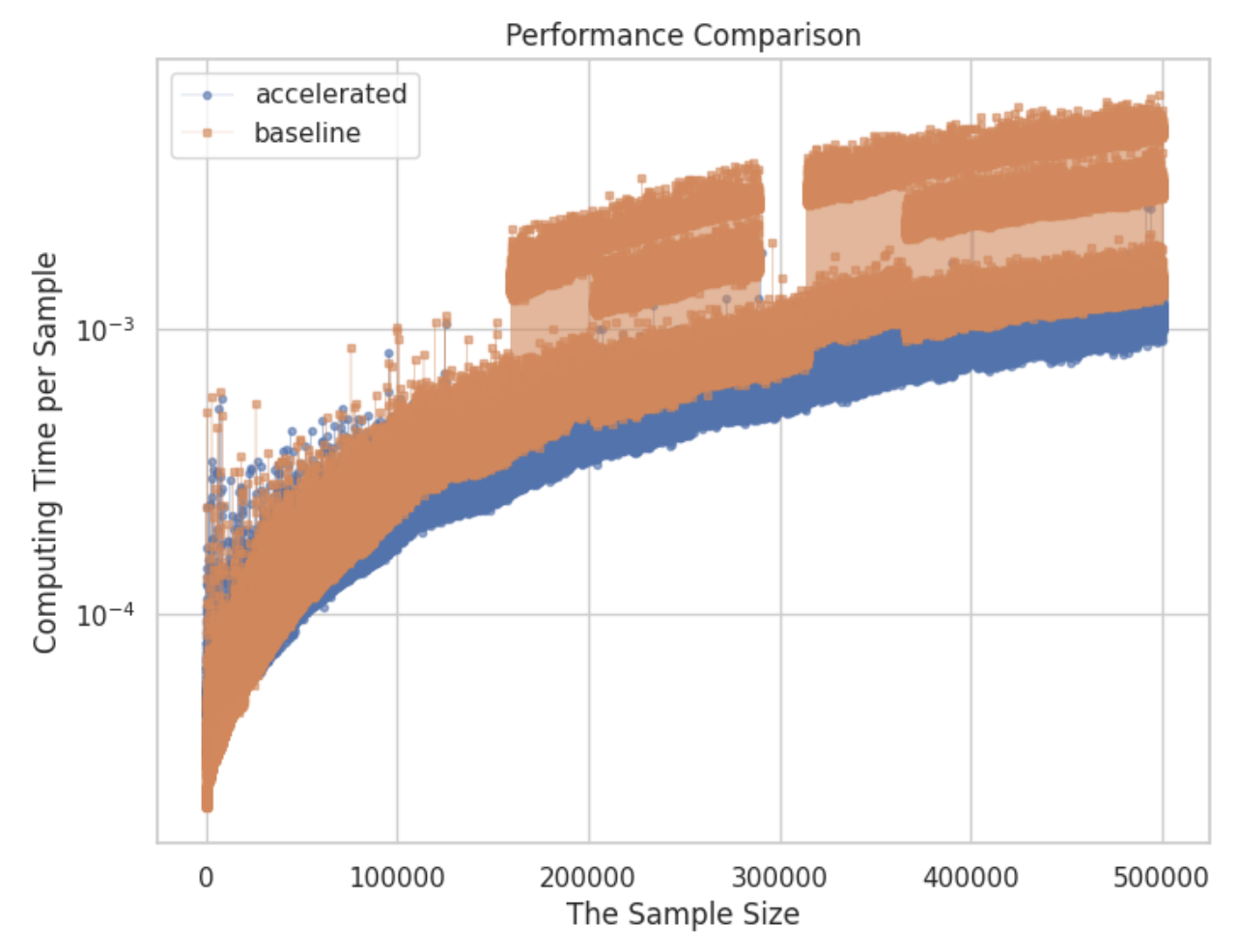}
    \caption{The performance of PKA}
    \label{fig:PKA}
\end{figure}

\begin{figure}[H]
    \centering
    \includegraphics[width=0.85\linewidth]{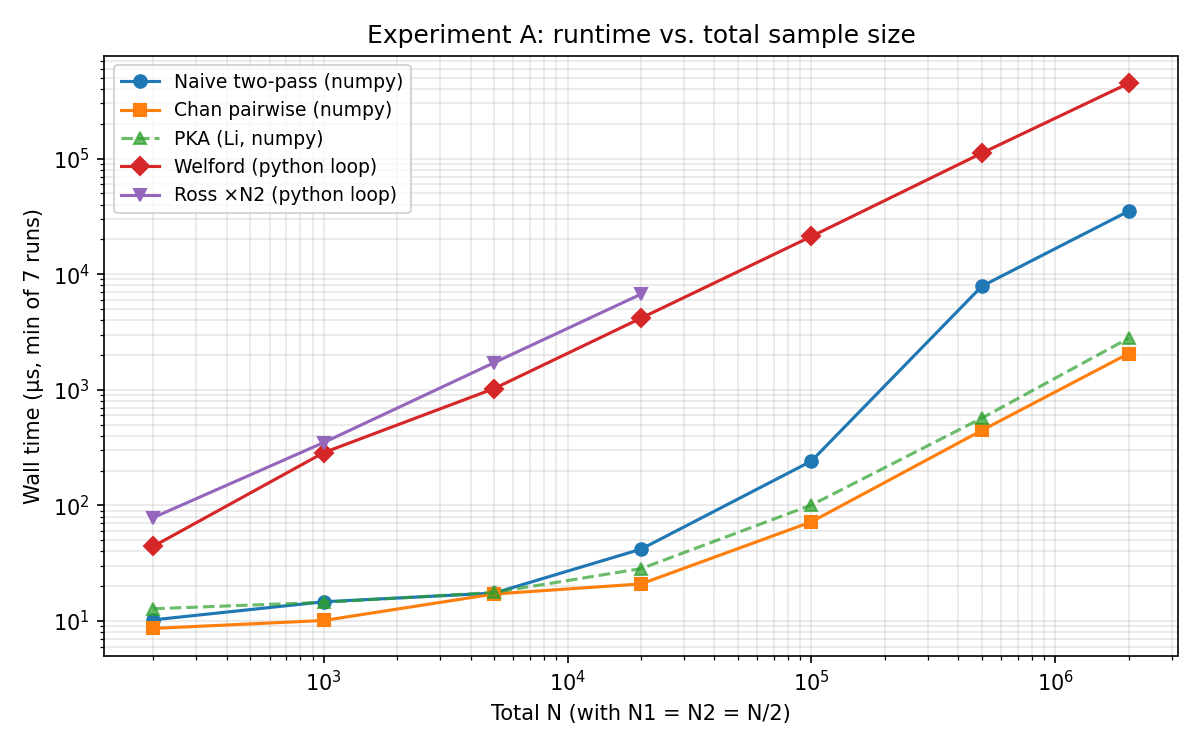}
    \caption{Runtime comparison across five methods 
    with $N_1 = N_2 = N/2$, using 
    \texttt{time.perf\_counter()} (minimum of 7 runs). 
    Chan pairwise and PKA (both NumPy-vectorised) dominate 
    for large $N$; Welford and Ross use per-sample Python loops.}
    \label{fig:runtime}
\end{figure}

As Fig.\ref{fig:PKA} shows, the PKA performance is the one labeled "accelerated" in the plot, which shows smaller computational time compared to baseline, and when the sample size gets larger this tendency becomes more apparent with $N_1=N_2$. When the total sample size N increases to 50,0000, the PKA reduces $22.04\%$ to $75.60\%$ computing times with statics confidence. This test is settled in the case that $N_1=N_2$, hence the paper creating another more precise experiment about it.

\begin{figure}[H]
    \centering
    \includegraphics[width=1\linewidth]{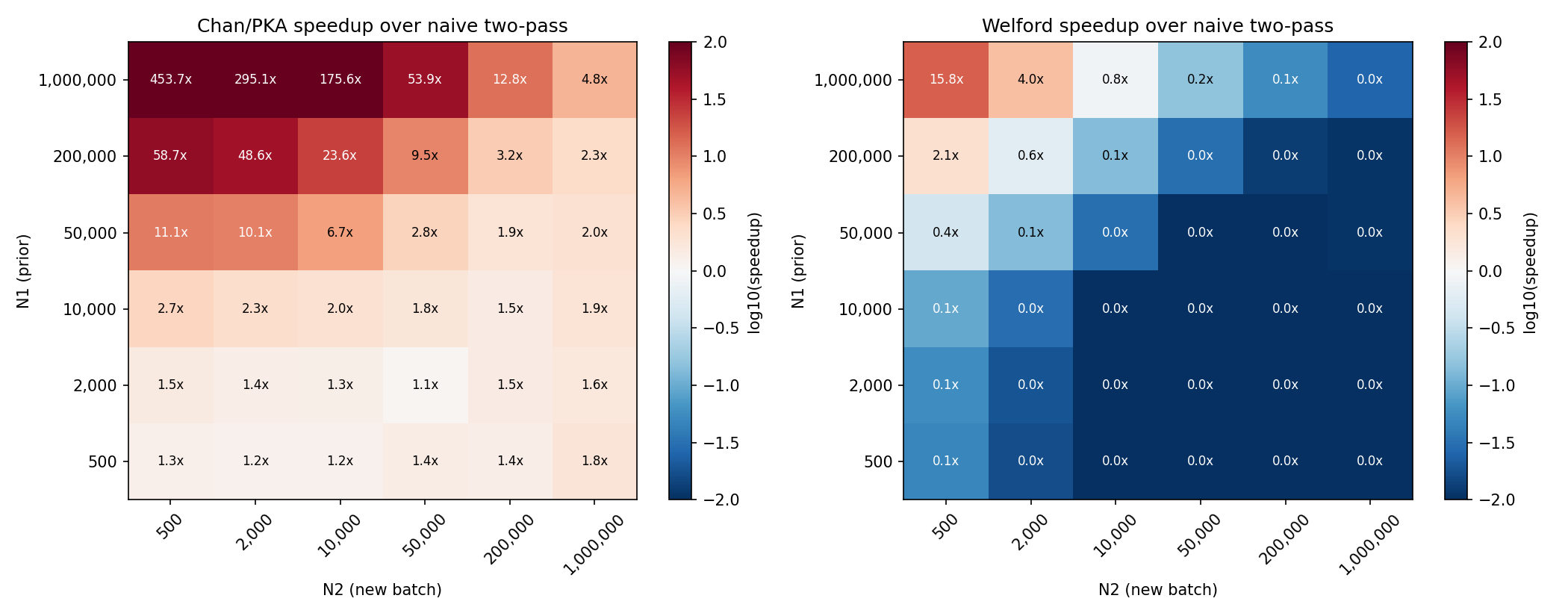}
    \caption{Speedup of Chan/PKA over na\"ive two-pass (left) 
    and Welford (right), varying $N_1$ and $N_2$. 
    Confirms that acceleration grows with $N_1/N_2$, 
    matching the theoretical prediction of 
    Eqs.~\eqref{eq:19}--\eqref{eq:20}.}
    \label{fig:heatmap}
\end{figure}

But we need a more clear view of the relationship of those variables, the paper making $N_1$ and $N_2$ range within $[10e3,10e5]$ for making the \textbf{second test}. And getting the mean values 30 times for smoothing the figure, the speedup heatmap confirms this trend in Fig.\ref{fig:heatmap}. According to the figure of the surface, the PKA starts to reduce the time of calculating variance roughly around $N_1=3e4$ until the end of testing values, indicating is effective when $N_1$ is larger than an unknown certain value.

The heatmap in Fig.~\ref{fig:heatmap} confirms that the speedup grows monotonically with $N_1/N_2$, consistent with the theoretical limits in Eqs.~\eqref{eq:19}--\eqref{eq:20}.

During the experiment, there is another thing that needs to be noticed, due to the Truncation Error \cite{error}, an error caused by limits of storage digits in the computer, there is no surprise that PKA holds a larger Truncation Error since it evolves more step to calculate as Fig.\ref{fig:graph} show, is on the order of the machine epsilon of the arithmetic type used ($\approx 1.2 \times 10^{-7}$ for \texttt{float32}, $\approx 2.2 \times 10^{-16}$ for \texttt{float64}). Since PKA involves more arithmetic operations than a single variance call, a moderately larger accumulated round-off is expected; we quantify this in our benchmarks.

To position PKA against established methods, we 
benchmark five algorithms under identical conditions 
(Fig.~\ref{fig:runtime}): na\"ive two-pass, Chan 
pairwise~\cite{classic}, PKA, Welford~\cite{Welford1962}, 
and Ross~\cite{ross} applied $N_2$ times. At 
$N = 2{,}000{,}000$ with $N_1 = N_2$, Chan pairwise 
completes in 2.1\,ms versus 35.2\,ms for na\"ive 
two-pass. PKA achieves 2.8\,ms. Welford's per-sample 
loop requires 449\,ms, confirming that batch-vectorised 
updates dominate per-sample iteration.

Fig.~\ref{fig:heatmap} shows the speedup across varying 
$(N_1, N_2)$ combinations. When $N_1 = 1{,}000{,}000$ 
and $N_2 = 500$, the speedup reaches $454\times$. This large factor is inherent to any incremental method in this regime; the key observation is that the measured speedup 
closely matches the theoretical $\tau_a$ predicted 
by Eqs.~\eqref{eq:19}--\eqref{eq:20}, confirming 
the validity of our constant-factor analysis.

\section{Real-time Experiment}
\label{sec:realtime}
The results in simulated tests were still lacking support when in an online environment, it might be more dynamic in the aspect of delays or computation power, hence the paper utilizes PKA to compute the variance of Individual Household Electric Power Consumption dataset \cite{UCI} from UCI Datasets to enhance the conclusion. The reason for choosing it is the form of the dataset is a stream record by day, as well as the easy accessibility to the data, making it perfect for testing the PKA method. In this experiment, we only calculate the feature \textbf{Global\_active\_power}. The research is using Python package \textbf{zmq} for achieving that(create two process bind to it), and combining the conditions that $S_1\times S_2=\{(N_{1_1},N_{2_1}),...|N_{1_1}\in S_1,N_{2_1}\in S_2\}$, which $S_1$ and $S_2$ simply are sets all equal to ${20,200,2000,20000,20000}$, for grid searching the results. The figure we get from the search still indicates the effectiveness of PKA, the larger the original dataset, the better the PKA works(the difference between the two plots,the PKA and baseline(calculate directly), gets larger).

\begin{figure}[H]
    \centering
    \includegraphics[width=0.9\linewidth]{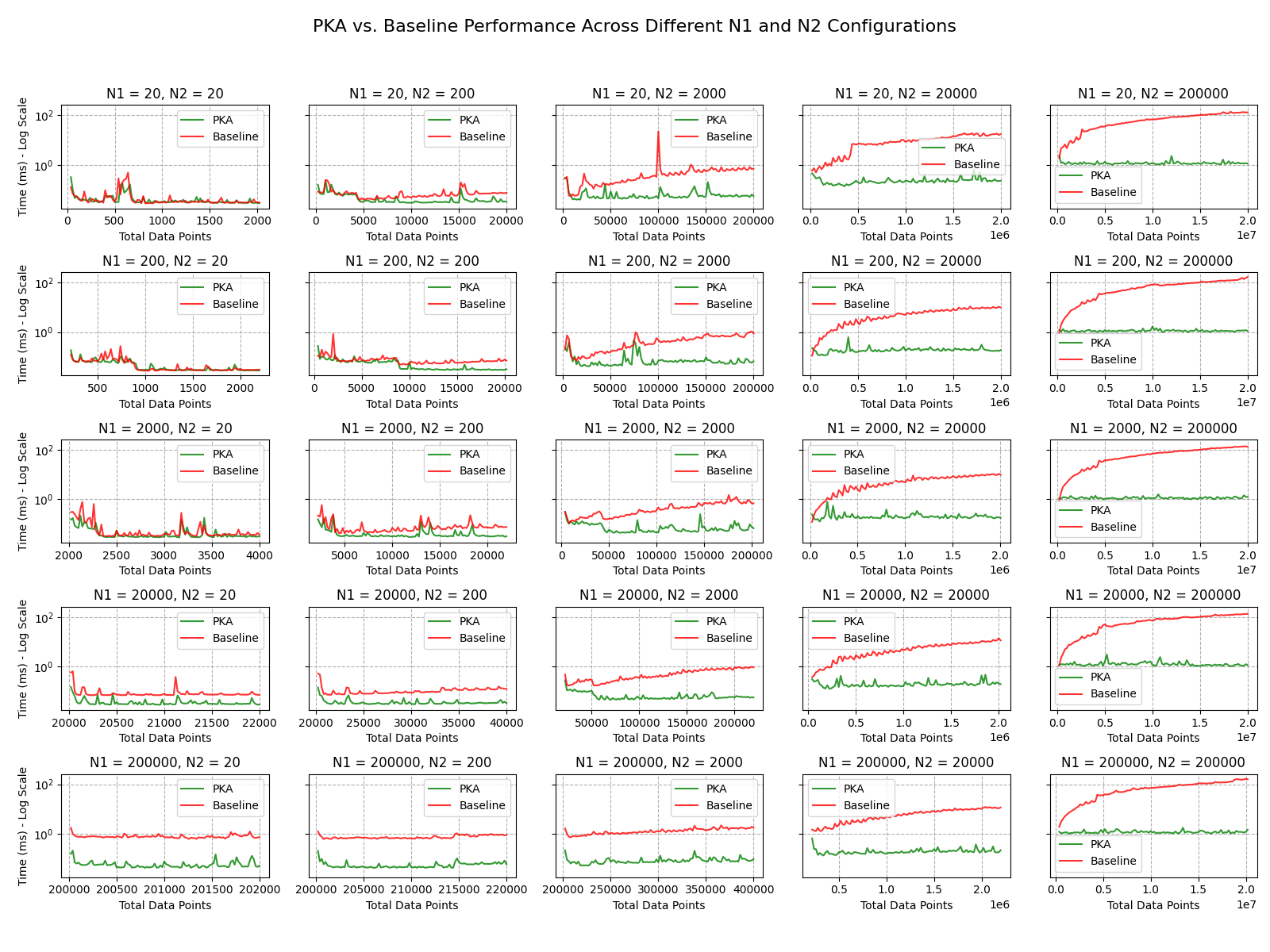}
    \caption{The comparison test in UCI power consumption dataset}
    \label{fig:real}
\end{figure}

\subsection{Numerical Stability and Memory Usage}

\begin{figure}[H]
    \centering
    \includegraphics[width=0.85\linewidth]{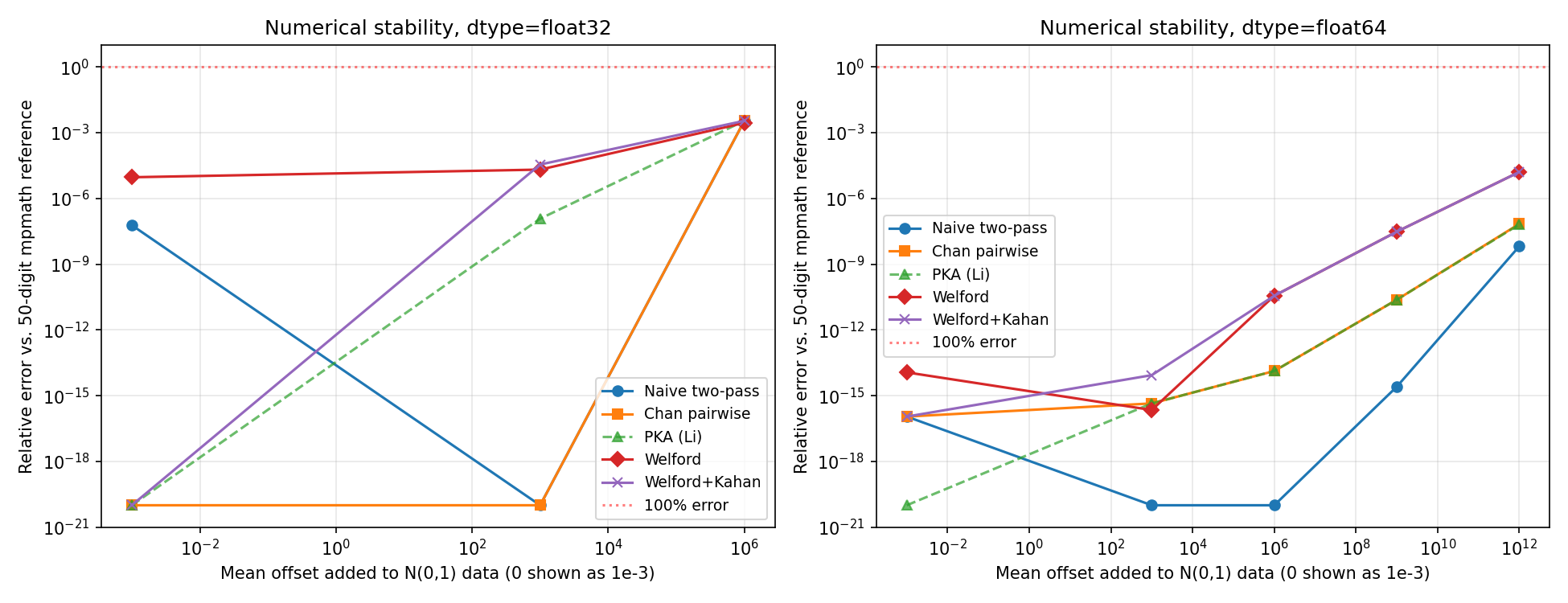}
    \caption{Relative error versus 50-digit reference 
    on ill-conditioned data (large mean offset added to 
    $\mathcal{N}(0,1)$ samples). Left: \texttt{float32}; 
    right: \texttt{float64}.}
    \label{fig:stability}
\end{figure}

To assess numerical stability, we construct ill-conditioned 
data by adding a large mean offset $M$ to $\mathcal{N}(0,1)$ samples, so the true variance remains~1 but values are of order~$M$. Fig.~\ref{fig:stability} compares the relative error of each method against a 50-digit \texttt{mpmath} reference. In \texttt{float64}, naive two-pass and Chan/PKA maintain accuracy up to offset $10^9$; Welford degrades earlier due to per-sample running-mean accumulation.

\begin{figure}[H]
    \centering
    \includegraphics[width=0.8\linewidth]{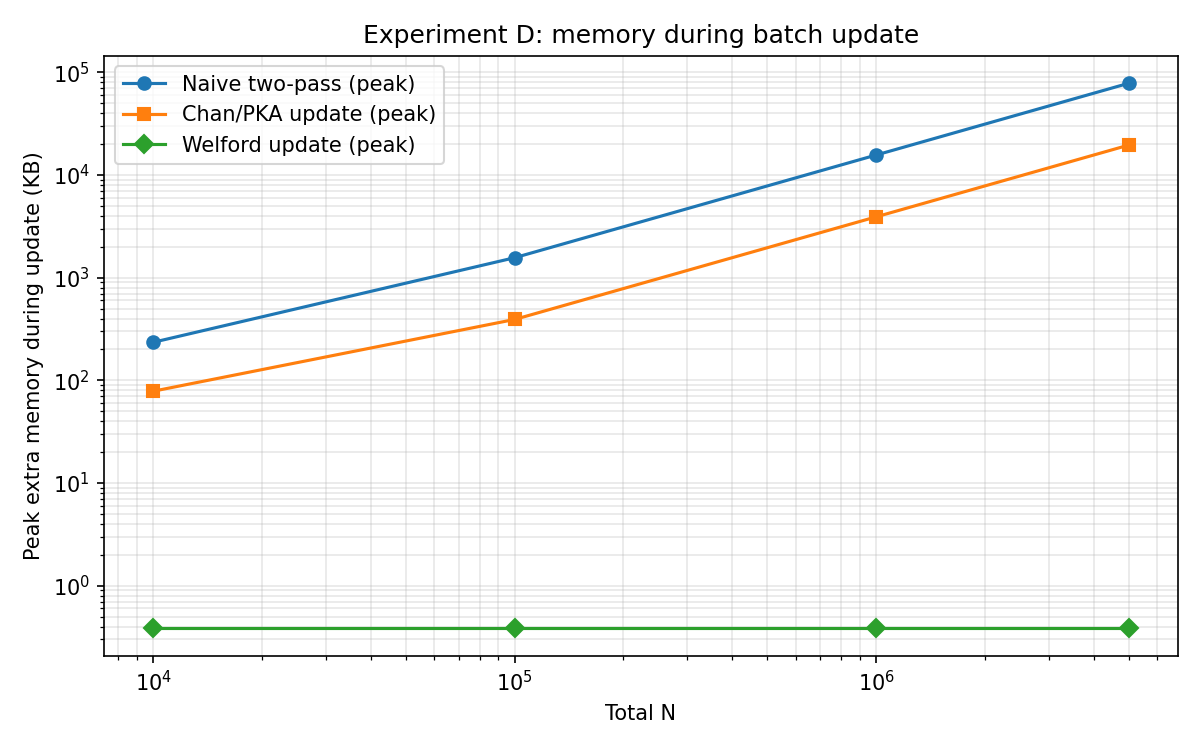}
    \caption{Peak memory during a batch update. 
    Welford uses $O(1)$; Chan/PKA uses $O(N_2)$ 
    temporary arrays; na\"ive two-pass uses $O(N)$.}
    \label{fig:memory}
\end{figure}

Fig.~\ref{fig:memory} shows peak memory during a batch 
update. Welford maintains constant $O(1)$ memory regardless 
of $N$; Chan/PKA requires $O(N_2)$ temporary arrays for 
vectorised computation; naive two-pass requires $O(N)$ since the full concatenated dataset must be materialised.

\section{Conclusion}
In general, the PKA method for calculating variance the article suggests holds an improvement in the computational burden of the variance calculations with the assistance of the knowledge of the variance of the original dataset, addressing critical challenges in large-scale and streaming data environments. When the factor $\tau_a$ fulfills the conditions, the PKA technique finds quicker computation when additional information is incorporated, making it beneficial, especially when processing large volumes of datasets during data analysis. Our theoretical analysis indicates that while PKA is highly effective when incorporating small to moderate-sized updates into a large dataset, its benefits diminish when the added data size approaches that of the original dataset.

Our findings also show those PKA methods can achieve reduction in computation time under general conditions. But the factor in all PKA methods are holding a potential issue when calculating variance, it assumes the unit time $u_m$ and $u_a$ are constants. The assumption of PKA in variance is approximately correct when processing large-size data, further improvements and validations to it may still needed in the future.Our benchmarks further show that PKA and Chan's pairwise formula yield comparable numerical stability to naive two-pass methods on well-conditioned data, while requiring significantly less memory than full recomputation (Figs.~\ref{fig:stability}--\ref{fig:memory}). Future work includes 
analysing PKA's effectiveness in dynamic environments 
and extending it to other metrics such as skewness.

\section*{Competing interest}
The authors declare that they have no known competing financial interests or personal relationships that could have appeared to influence the work reported in this paper.

\section*{Funding Information}
Not Applicable

\section*{Author Contribution}
Not Applicable

\section*{Data Availability Statement}  
The data were generated using \texttt{np.random.randn} with a fixed seed (8086) for reproducibility. Upon publication, the experiment will be available at \url{https://www.kaggle.com/code/spike8086/pka-method} and \url{https://www.kaggle.com/code/spike8086/pka-real-time-test}.  

\section*{Research Involving Human or Animals}
Not Applicable

\section*{Informed Consent}
Not applicable.

\bibliography{sn-article.bbl}

\begin{thebibliography}{}
\renewcommand{\doi}[1]{\url{https://doi.org/#1}}
\bibcommenthead

\bibitem [\protect \citeauthoryear {%
Agterberg%
}{%
Agterberg%
}{%
{\protect \APACyear {1993}}%
}]{%
cov}
\APACinsertmetastar {%
cov}%
\begin{APACrefauthors}%
Agterberg, F.P.%
\end{APACrefauthors}%
\unskip\
\newblock
\APACrefYearMonthDay{1993}{{\APACmonth{12}}}{}.
\newblock
{\BBOQ}\APACrefatitle {Calculation of the variance of mean values for blocks in regional resource evaluation studies} {Calculation of the variance of mean values for blocks in regional resource evaluation studies}.{\BBCQ}
\newblock
\APACjournalVolNumPages{Nonrenewable Resources}{2}{4}{312–324,}
\newblock
\begin{APACrefDOI} \doi{10.1007/bf02257541} \end{APACrefDOI}
\newblock

\newblock

\PrintBackRefs{\CurrentBib}

\bibitem [\protect \citeauthoryear {%
Bekci%
}{%
Bekci%
}{%
{\protect \APACyear {2024}}%
}]{%
issue2}
\APACinsertmetastar {%
issue2}%
\begin{APACrefauthors}%
Bekci, R.Y.%
\end{APACrefauthors}%
\unskip\
\newblock
\APACrefYearMonthDay{2024}{}{}.
\newblock
{\BBOQ}\APACrefatitle {Online Learning of Delayed Choices} {Online learning of delayed choices}.{\BBCQ}
\newblock
 A.~Globerson\ \BOthers {.}\ (\BEDS), \APACrefbtitle {Advances in Neural Information Processing Systems} {Advances in neural information processing systems}\ (\BVOL~37, \BPGS\ 2292--2322).
\newblock
\APACaddressPublisher{}{Curran Associates, Inc.}
\PrintBackRefs{\CurrentBib}

\bibitem [\protect \citeauthoryear {%
Bennett%
, Grout%
, P{\'e}bay%
, Roe%
\BCBL {}\ \BBA {} Thompson%
}{%
Bennett%
\ \protect \BOthers {.}}{%
{\protect \APACyear {2009}}%
}]{%
Bennett2009}
\APACinsertmetastar {%
Bennett2009}%
\begin{APACrefauthors}%
Bennett, J.%
, Grout, R.%
, P{\'e}bay, P.%
, Roe, D.%
\BCBL {} Thompson, D.%
\end{APACrefauthors}%
\unskip\
\newblock
\APACrefYearMonthDay{2009}{}{}.
\newblock
{\BBOQ}\APACrefatitle {Numerically Stable, Single-Pass, Parallel Statistics Algorithms} {Numerically stable, single-pass, parallel statistics algorithms}.{\BBCQ}
\newblock
 \APACrefbtitle {IEEE International Conference on Cluster Computing (CLUSTER)} {Ieee international conference on cluster computing (cluster)}\ (\BPGS\ 1--8).
\PrintBackRefs{\CurrentBib}

\bibitem [\protect \citeauthoryear {%
Chan%
, Golub%
\BCBL {}\ \BBA {} LeVeque%
}{%
Chan%
\ \protect \BOthers {.}}{%
{\protect \APACyear {1983}}%
}]{%
classic}
\APACinsertmetastar {%
classic}%
\begin{APACrefauthors}%
Chan, T.F.%
, Golub, G.H.%
\BCBL {} LeVeque, R.J.%
\end{APACrefauthors}%
\unskip\
\newblock
\APACrefYearMonthDay{1983}{}{}.
\newblock
{\BBOQ}\APACrefatitle {Algorithms for computing the sample variance: Analysis and recommendations} {Algorithms for computing the sample variance: Analysis and recommendations}.{\BBCQ}
\newblock
\APACjournalVolNumPages{The American Statistician}{37}{3}{242--247,}
\newblock
\begin{APACrefDOI} \doi{10.1080/00031305.1983.10483115} \end{APACrefDOI}
\newblock

\newblock

\PrintBackRefs{\CurrentBib}

\bibitem [\protect \citeauthoryear {%
Dempster%
, Laird%
\BCBL {}\ \BBA {} Rubin%
}{%
Dempster%
\ \protect \BOthers {.}}{%
{\protect \APACyear {1977}}%
}]{%
EM}
\APACinsertmetastar {%
EM}%
\begin{APACrefauthors}%
Dempster, A.P.%
, Laird, N.M.%
\BCBL {} Rubin, D.B.%
\end{APACrefauthors}%
\unskip\
\newblock
\APACrefYearMonthDay{1977}{{\APACmonth{09}}}{}.
\newblock
{\BBOQ}\APACrefatitle {Maximum Likelihood from Incomplete Data Via the EM Algorithm} {Maximum likelihood from incomplete data via the em algorithm}.{\BBCQ}
\newblock
\APACjournalVolNumPages{Journal of the Royal Statistical Society Series B: Statistical Methodology}{39}{1}{1–22,}
\newblock
\begin{APACrefURL} {http://dx.doi.org/10.1111/j.2517-6161.1977.tb01600.x} \end{APACrefURL}
\newblock

\newblock

\PrintBackRefs{\CurrentBib}

\bibitem [\protect \citeauthoryear {%
Fisher%
}{%
Fisher%
}{%
{\protect \APACyear {1935}}%
}]{%
fisher}
\APACinsertmetastar {%
fisher}%
\begin{APACrefauthors}%
Fisher, R.A.%
\end{APACrefauthors}%
\unskip\
\newblock
\APACrefYear{1935}.
\newblock
\APACrefbtitle {{The Design of Experiments}} {{The Design of Experiments}}.
\newblock
\APACaddressPublisher{Edinburgh}{Oliver and Boyd}.
\PrintBackRefs{\CurrentBib}

\bibitem [\protect \citeauthoryear {%
Guan%
\ \BBA {} Burton%
}{%
Guan%
\ \BBA {} Burton%
}{%
{\protect \APACyear {2022}}%
}]{%
ensemble}
\APACinsertmetastar {%
ensemble}%
\begin{APACrefauthors}%
Guan, X.%
\BCBT {}\ \BBA {} Burton, H.%
\end{APACrefauthors}%
\unskip\
\newblock
\APACrefYearMonthDay{2022}{}{}.
\newblock
{\BBOQ}\APACrefatitle {Bias-variance tradeoff in machine learning: Theoretical formulation and implications to structural engineering applications} {Bias-variance tradeoff in machine learning: Theoretical formulation and implications to structural engineering applications}.{\BBCQ}
\newblock
\APACjournalVolNumPages{Structures}{46}{}{17-30,}
\newblock
\begin{APACrefDOI} \doi{10.1016/j.istruc.2022.10.004} \end{APACrefDOI}
\newblock

\newblock

\PrintBackRefs{\CurrentBib}

\bibitem [\protect \citeauthoryear {%
Hanson%
}{%
Hanson%
}{%
{\protect \APACyear {1975}}%
}]{%
Hanson1975}
\APACinsertmetastar {%
Hanson1975}%
\begin{APACrefauthors}%
Hanson, R.J.%
\end{APACrefauthors}%
\unskip\
\newblock
\APACrefYearMonthDay{1975}{}{}.
\newblock
{\BBOQ}\APACrefatitle {Stably Updating Mean and Standard Deviation of Data} {Stably updating mean and standard deviation of data}.{\BBCQ}
\newblock
\APACjournalVolNumPages{Communications of the ACM}{18}{1}{57--58,}
\newblock

\newblock

\PrintBackRefs{\CurrentBib}

\bibitem [\protect \citeauthoryear {%
Hastie%
, Tibshirani%
\BCBL {}\ \BBA {} Friedman%
}{%
Hastie%
\ \protect \BOthers {.}}{%
{\protect \APACyear {2009}}%
}]{%
gaussian}
\APACinsertmetastar {%
gaussian}%
\begin{APACrefauthors}%
Hastie, T.%
, Tibshirani, R.%
\BCBL {} Friedman, J.%
\end{APACrefauthors}%
\unskip\
\newblock
\APACrefYear{2009}.
\newblock
\APACrefbtitle {The elements of statistical learning} {The elements of statistical learning}.
\newblock
\APACaddressPublisher{}{Springer New York}.
\PrintBackRefs{\CurrentBib}

\bibitem [\protect \citeauthoryear {%
Hebrail%
\ \BBA {} Berard%
}{%
Hebrail%
\ \BBA {} Berard%
}{%
{\protect \APACyear {2006}}%
}]{%
UCI}
\APACinsertmetastar {%
UCI}%
\begin{APACrefauthors}%
Hebrail, G.%
\BCBT {}\ \BBA {} Berard, A.%
\end{APACrefauthors}%
\unskip\
\newblock
\APACrefYearMonthDay{2006}{}{}.
\newblock
\APACrefbtitle {{Individual Household Electric Power Consumption}.} {{Individual Household Electric Power Consumption}.}
\newblock
\APAChowpublished {UCI Machine Learning Repository}.
\newblock
\APACrefnote{{DOI}: https://doi.org/10.24432/C58K54}
\PrintBackRefs{\CurrentBib}

\bibitem [\protect \citeauthoryear {%
Hu%
, Li%
\BCBL {}\ \BBA {} Shi%
}{%
Hu%
\ \protect \BOthers {.}}{%
{\protect \APACyear {2023}}%
}]{%
issue1}
\APACinsertmetastar {%
issue1}%
\begin{APACrefauthors}%
Hu, S.%
, Li, G.%
\BCBL {} Shi, W.%
\end{APACrefauthors}%
\unskip\
\newblock
\APACrefYearMonthDay{2023}{}{}.
\newblock
{\BBOQ}\APACrefatitle {LARS: A Latency-Aware and Real-Time Scheduling Framework for Edge-Enabled Internet of Vehicles} {Lars: A latency-aware and real-time scheduling framework for edge-enabled internet of vehicles}.{\BBCQ}
\newblock
\APACjournalVolNumPages{IEEE Transactions on Services Computing}{16}{1}{398-411,}
\newblock
\begin{APACrefDOI} \doi{10.1109/TSC.2021.3106260} \end{APACrefDOI}
\newblock

\newblock

\PrintBackRefs{\CurrentBib}

\bibitem [\protect \citeauthoryear {%
Knuth%
}{%
Knuth%
}{%
{\protect \APACyear {2005}}%
}]{%
error}
\APACinsertmetastar {%
error}%
\begin{APACrefauthors}%
Knuth, D.E.%
\end{APACrefauthors}%
\unskip\
\newblock
\APACrefYear{2005}.
\newblock
\APACrefbtitle {The Art of Computer Programming, Volume 1, Fascicle 1: MMIX A RISC Computer for the New Millenium} {The art of computer programming, volume 1, fascicle 1: Mmix a risc computer for the new millenium}.
\newblock
\APACaddressPublisher{}{Addison-Wesley}.
\PrintBackRefs{\CurrentBib}

\bibitem [\protect \citeauthoryear {%
MacQueen%
}{%
MacQueen%
}{%
{\protect \APACyear {1967}}%
}]{%
WCSS}
\APACinsertmetastar {%
WCSS}%
\begin{APACrefauthors}%
MacQueen, J.%
\end{APACrefauthors}%
\unskip\
\newblock
\APACrefYearMonthDay{1967}{}{}.
\newblock
{\BBOQ}\APACrefatitle {Some methods for classification and analysis of multivariate observations} {Some methods for classification and analysis of multivariate observations}.{\BBCQ}.
\newblock
\begin{APACrefURL} {https://api.semanticscholar.org/CorpusID:6278891} \end{APACrefURL}
\PrintBackRefs{\CurrentBib}

\bibitem [\protect \citeauthoryear {%
\APACciteatitle {Online algorithm for variance components estimation}}{%
\APACciteatitle {Online algorithm for variance components estimation}}{%
{\protect \APACyear {2021}}%
}]{%
attempts}
\APACinsertmetastar {%
attempts}%
{\BBOQ}\APACrefatitle {Online algorithm for variance components estimation} {Online algorithm for variance components estimation}.{\BBCQ}
\newblock
\APACrefYearMonthDay{2021}{}{}.
\newblock
\APACjournalVolNumPages{Communications in Nonlinear Science and Numerical Simulation}{97}{}{105722,}
\newblock
\begin{APACrefDOI} \doi{10.1016/j.cnsns.2021.105722} \end{APACrefDOI}
\newblock

\newblock

\PrintBackRefs{\CurrentBib}

\bibitem [\protect \citeauthoryear {%
P.~P{\'e}bay%
, Terriberry%
, Kolla%
\BCBL {}\ \BBA {} Bennett%
}{%
P.~P{\'e}bay%
\ \protect \BOthers {.}}{%
{\protect \APACyear {2016}}%
}]{%
Pebay2016}
\APACinsertmetastar {%
Pebay2016}%
\begin{APACrefauthors}%
P{\'e}bay, P.%
, Terriberry, T.B.%
, Kolla, H.%
\BCBL {} Bennett, J.%
\end{APACrefauthors}%
\unskip\
\newblock
\APACrefYearMonthDay{2016}{}{}.
\newblock
{\BBOQ}\APACrefatitle {Numerically Stable, Scalable Formulas for Parallel and Online Computation of Higher-Order Multivariate Central Moments with Arbitrary Weights} {Numerically stable, scalable formulas for parallel and online computation of higher-order multivariate central moments with arbitrary weights}.{\BBCQ}
\newblock
\APACjournalVolNumPages{Computational Statistics}{31}{4}{1305--1325,}
\newblock
\begin{APACrefDOI} \doi{10.1007/s00180-015-0637-z} \end{APACrefDOI}
\newblock

\newblock

\PrintBackRefs{\CurrentBib}

\bibitem [\protect \citeauthoryear {%
P.P.~P{\'e}bay%
}{%
P.P.~P{\'e}bay%
}{%
{\protect \APACyear {2008}}%
}]{%
Pebay2008}
\APACinsertmetastar {%
Pebay2008}%
\begin{APACrefauthors}%
P{\'e}bay, P.P.%
\end{APACrefauthors}%
\unskip\
\newblock
\APACrefYearMonthDay{2008}{}{}.
\newblock
\APACrefbtitle {Formulas for Robust, One-Pass Parallel Computation of Covariances and Arbitrary-Order Statistical Moments} {Formulas for robust, one-pass parallel computation of covariances and arbitrary-order statistical moments}\ \APACbVolEdTR{}{\BTR{}\ \BNUM\ SAND2008-6212}.
\newblock
\APACaddressInstitution{}{Sandia National Laboratories}.
\PrintBackRefs{\CurrentBib}

\bibitem [\protect \citeauthoryear {%
Ross%
}{%
Ross%
}{%
{\protect \APACyear {2021}}%
}]{%
ross}
\APACinsertmetastar {%
ross}%
\begin{APACrefauthors}%
Ross, S.M.%
\end{APACrefauthors}%
\unskip\
\newblock
\APACrefYear{2021}.
\newblock
\APACrefbtitle {Introduction to probability and statistics for engineers and scientists} {Introduction to probability and statistics for engineers and scientists}.
\newblock
\APACaddressPublisher{London, United Kingdom}{Academic Press}.
\PrintBackRefs{\CurrentBib}

\bibitem [\protect \citeauthoryear {%
Schmitt%
\ \BBA {} Fessler%
}{%
Schmitt%
\ \BBA {} Fessler%
}{%
{\protect \APACyear {2012}}%
}]{%
CT-Var}
\APACinsertmetastar {%
CT-Var}%
\begin{APACrefauthors}%
Schmitt, S.M.%
\BCBT {}\ \BBA {} Fessler, J.A.%
\end{APACrefauthors}%
\unskip\
\newblock
\APACrefYearMonthDay{2012}{}{}.
\newblock
{\BBOQ}\APACrefatitle {Fast variance computation for quadratically penalized iterative reconstruction of 3D axial CT images} {Fast variance computation for quadratically penalized iterative reconstruction of 3d axial ct images}.{\BBCQ}
\newblock
 \APACrefbtitle {2012 IEEE Nuclear Science Symposium and Medical Imaging Conference Record (NSS/MIC)} {2012 ieee nuclear science symposium and medical imaging conference record (nss/mic)}\ (\BPG~3287-3292).
\PrintBackRefs{\CurrentBib}

\bibitem [\protect \citeauthoryear {%
Schubert%
\ \BBA {} Gertz%
}{%
Schubert%
\ \BBA {} Gertz%
}{%
{\protect \APACyear {2018}}%
}]{%
Schubert2018}
\APACinsertmetastar {%
Schubert2018}%
\begin{APACrefauthors}%
Schubert, E.%
\BCBT {}\ \BBA {} Gertz, M.%
\end{APACrefauthors}%
\unskip\
\newblock
\APACrefYearMonthDay{2018}{}{}.
\newblock
{\BBOQ}\APACrefatitle {Numerically Stable Parallel Computation of (Co-)Variance} {Numerically stable parallel computation of (co-)variance}.{\BBCQ}
\newblock
 \APACrefbtitle {Proceedings of the 30th International Conference on Scientific and Statistical Database Management (SSDBM)} {Proceedings of the 30th international conference on scientific and statistical database management (ssdbm)}\ (\BPGS\ 10:1--10:12).
\newblock
\APACrefnote{SSDBM 2018 Best Paper Award}
\PrintBackRefs{\CurrentBib}

\bibitem [\protect \citeauthoryear {%
Searcóid%
}{%
Searcóid%
}{%
{\protect \APACyear {2006}}%
}]{%
lip}
\APACinsertmetastar {%
lip}%
\begin{APACrefauthors}%
Searcóid, M.%
\end{APACrefauthors}%
\unskip\
\newblock
\APACrefYear{2006}.
\newblock
\APACrefbtitle {Lipschitz Functions} {Lipschitz functions}.
\newblock
\APACaddressPublisher{Berlin, New York}{Springer-Verlag}.
\PrintBackRefs{\CurrentBib}

\bibitem [\protect \citeauthoryear {%
Shalev-Shwartz%
}{%
Shalev-Shwartz%
}{%
{\protect \APACyear {2012}}%
}]{%
online}
\APACinsertmetastar {%
online}%
\begin{APACrefauthors}%
Shalev-Shwartz, S.%
\end{APACrefauthors}%
\unskip\
\newblock
\APACrefYearMonthDay{2012}{}{}.
\newblock
{\BBOQ}\APACrefatitle {Online Learning and Online Convex Optimization} {Online learning and online convex optimization}.{\BBCQ}
\newblock
\APACjournalVolNumPages{Foundations and Trends in Machine Learning}{4}{2}{107--194,}
\newblock
\begin{APACrefDOI} \doi{10.1561/2200000018} \end{APACrefDOI}
\newblock

\newblock

\PrintBackRefs{\CurrentBib}

\bibitem [\protect \citeauthoryear {%
Welford%
}{%
Welford%
}{%
{\protect \APACyear {1962}}%
}]{%
Welford1962}
\APACinsertmetastar {%
Welford1962}%
\begin{APACrefauthors}%
Welford, B.P.%
\end{APACrefauthors}%
\unskip\
\newblock
\APACrefYearMonthDay{1962}{}{}.
\newblock
{\BBOQ}\APACrefatitle {Note on a Method for Calculating Corrected Sums of Squares and Products} {Note on a method for calculating corrected sums of squares and products}.{\BBCQ}
\newblock
\APACjournalVolNumPages{Technometrics}{4}{3}{419--420,}
\newblock
\begin{APACrefDOI} \doi{10.1080/00401706.1962.10490022} \end{APACrefDOI}
\newblock

\newblock

\PrintBackRefs{\CurrentBib}

\bibitem [\protect \citeauthoryear {%
West%
}{%
West%
}{%
{\protect \APACyear {1979}}%
}]{%
West1979}
\APACinsertmetastar {%
West1979}%
\begin{APACrefauthors}%
West, D.H.D.%
\end{APACrefauthors}%
\unskip\
\newblock
\APACrefYearMonthDay{1979}{}{}.
\newblock
{\BBOQ}\APACrefatitle {Updating Mean and Variance Estimates: An Improved Method} {Updating mean and variance estimates: An improved method}.{\BBCQ}
\newblock
\APACjournalVolNumPages{Communications of the ACM}{22}{9}{532--535,}
\newblock

\newblock

\PrintBackRefs{\CurrentBib}

\bibitem [\protect \citeauthoryear {%
Youngs%
\ \BBA {} Cramer%
}{%
Youngs%
\ \BBA {} Cramer%
}{%
{\protect \APACyear {1971}}%
}]{%
Youngs1971}
\APACinsertmetastar {%
Youngs1971}%
\begin{APACrefauthors}%
Youngs, E.A.%
\BCBT {}\ \BBA {} Cramer, E.M.%
\end{APACrefauthors}%
\unskip\
\newblock
\APACrefYearMonthDay{1971}{}{}.
\newblock
{\BBOQ}\APACrefatitle {Some Results Relevant to Choice of Sum and Sum-of-Product Algorithms} {Some results relevant to choice of sum and sum-of-product algorithms}.{\BBCQ}
\newblock
\APACjournalVolNumPages{Technometrics}{13}{3}{657--665,}
\newblock
\begin{APACrefDOI} \doi{10.1080/00401706.1971.10488826} \end{APACrefDOI}
\newblock

\newblock

\PrintBackRefs{\CurrentBib}

\bibitem [\protect \citeauthoryear {%
Zhai%
, Wang%
\BCBL {}\ \BBA {} Wang%
}{%
Zhai%
\ \protect \BOthers {.}}{%
{\protect \APACyear {2014}}%
}]{%
ELM}
\APACinsertmetastar {%
ELM}%
\begin{APACrefauthors}%
Zhai, J.%
, Wang, J.%
\BCBL {} Wang, X.%
\end{APACrefauthors}%
\unskip\
\newblock
\APACrefYearMonthDay{2014}{}{}.
\newblock
{\BBOQ}\APACrefatitle {Ensemble online sequential extreme learning machine for large data set classification} {Ensemble online sequential extreme learning machine for large data set classification}.{\BBCQ}
\newblock
 \APACrefbtitle {2014 IEEE International Conference on Systems, Man, and Cybernetics (SMC)} {2014 ieee international conference on systems, man, and cybernetics (smc)}\ (\BPG~2250-2255).
\PrintBackRefs{\CurrentBib}

\bibitem [\protect \citeauthoryear {%
Zhang%
\ \BBA {} Lu%
}{%
Zhang%
\ \BBA {} Lu%
}{%
{\protect \APACyear {2021}}%
}]{%
promising}
\APACinsertmetastar {%
promising}%
\begin{APACrefauthors}%
Zhang, X.%
\BCBT {}\ \BBA {} Lu, X.%
\end{APACrefauthors}%
\unskip\
\newblock
\APACrefYearMonthDay{2021}{}{}.
\newblock
{\BBOQ}\APACrefatitle {Online algorithm for variance components estimation} {Online algorithm for variance components estimation}.{\BBCQ}
\newblock
\APACjournalVolNumPages{Communications in Nonlinear Science and Numerical Simulation}{97}{}{105722,}
\newblock
\begin{APACrefDOI} \doi{https://doi.org/10.1016/j.cnsns.2021.105722} \end{APACrefDOI}
\newblock

\newblock

\PrintBackRefs{\CurrentBib}

\end{thebibliography}

\end{document}